\documentclass{article}
\usepackage{PRIMEarxiv}

\usepackage{graphicx}

\usepackage{eucal} 

\usepackage{hyperref}
\usepackage{float}
\usepackage{verbatim} 
\restylefloat{figure}
\restylefloat{table}

\usepackage{tabularx,booktabs}
\usepackage{amsmath}
\usepackage{amsfonts} 
\usepackage[algo2e,ruled,lined, linesnumbered, nofillcomment, noend]{algorithm2e}
\usepackage{makecell}

\begin{document}










\pagestyle{fancy}
\thispagestyle{empty}
\rhead{ \textit{ }} 
\fancyhead[LO]{Efficient pattern-based anomaly detection in a network of multivariate devices}

\title{Efficient pattern-based anomaly detection in a network of multivariate devices}





\author{
  Len Feremans \\
  Department of Computer Science \\
  University of Antwerp\\
  Belgium\\
  \texttt{len.feremans@uantwerpen.be} \\
   \And
  Boris Cule \\
  Department of Cognitive Science and Artificial Intelligence\\
  Tilburg University \\
  Netherlands\\
  \texttt{b.cule@tilburguniversity.edu} \\
   \And
  Bart Goethals\\
  Department of Computer Science \\
  University of Antwerp\\
  Belgium\\
  \texttt{bart.goethals@uantwerpen.be} \\
}

\maketitle

\begin{abstract}
Many organisations manage service quality and monitor a large set devices and servers where each entity is associated with telemetry or physical sensor data series. Recently, various methods have been proposed to detect behavioural anomalies, however existing approaches focus on multivariate time series and ignore communication between entities. Moreover, we aim to support end-users in not only in locating  entities and sensors causing an anomaly at a certain period, but also explain this decision. 
We propose a scalable approach to detect anomalies using a two-step approach. First, we recover relations between entities in the network,  since relations are often dynamic in nature and caused by an unknown underlying process. Next, we report anomalies based on an embedding of sequential patterns. Pattern mining is efficient and supports interpretation, i.e. patterns represent frequent occurring behaviour in time series. We extend pattern mining to filter sequential patterns based on frequency, temporal constraints and minimum description length.
We collect and release two public datasets for international broadcasting and X from an Internet company. \textit{BAD} achieves an overall F1-Score of 0.78 on 9 benchmark datasets, significantly outperforming the best baseline by 3\%. Additionally, \textit{BAD} is also an order-of-magnitude faster than state-of-the-art anomaly detection methods.

\end{abstract}



\section{Introduction}
\label{intro}


A large stream of information is produced by networks of connected devices in the current Internet of Things world. 
Many companies are recording temporal data streams. However, given a large network of devices it becomes infeasible to inspect sensor logs manually and traditional expert-based supervision methods are prone to false positives where errors are often detected late or not at all~\cite{audibert2020usad}. Anomaly detection is an active research topic with applications in event logs, 
time series, multivariate data and networks \cite{lavin2015evaluating,liu2008isolation,savage2014anomaly,su2019robust}. 
We focus on \emph{contextual} anomaly detection in devices consisting of multivariate time series 
where operators are interested in detecting if a device behaves unexpectedly during a time period \cite{chandola2009anomaly}. Since it is difficult to collect labels we study this as an \emph{unsupervised} problem.
%
Our research is motivated by datasets collected from a company that monitors a \emph{heterogeneous network} of different devices each consisting of several sensors which we make publicly available. 
It occurs often that data is sent to multiple devices and it is expected that because of load-balancing and broadcasting connected entities exhibit anomalies at the same time. 

Recent methods such as \textsc{OmniAnomaly} and \textsc{Usad} detect behavioural anomalies in a single
entity consisting of multiple univariate time series based
on recurrent neural networks and autoencoders ~\cite{su2019robust,audibert2020usad,chen2022net}.
However these methods ignores the network context.
Related work, that takes the network into account, is  studied specific to a single domain, such as social networks \cite{savage2014anomaly} or intrusion detection \cite{xie2011anomaly}. 


We argue that including additional information for an entity in a network, such as time series of connected entities, leads to an increase in anomaly detection accuracy. 
For identifying intra-device similarities, we compare with algorithms (and representations) that support fast similarity search in time series databases. Piecewise Aggregate Approximation (\textsc{Paa}) and Product Quantization reduce time series dimensions and scale to billions of time series using GPU processing \cite{keogh2001dimensionality,jegou2010product}.      
For \emph{heterogeneous times databases}, \emph{feature-based} similarity measures have been proposed  \cite{hyndman2015large,bandara2020forecasting} where time series of different types are clustered based on multiple aspects such as the distribution, peaks, entropy, change points and patterns, instead of computing the similarity of point values using a distance metric.  
We focus on an efficient algorithm using \emph{feature-based} similarity to tackle the high variety in time series types. Our similarity score consider features based on the  overall trend, peaks and the distribution of a time series.  After construction the similarity matrix we could apply \emph{domain knowledge} to filter spurious relations. However, domain knowledge related to sensor types is generally not available.  Hence, we reverse engineer domain knowledge thereby computing patterns or \emph{frequent relation types}, to further spurious relations.  We also propose the usage of \emph{density-based fingerprints}, which is a novel approximation of the \emph{maximal information coefficient}~\cite{kinney2014equitability}.


 
\begin{figure}[b!]
    \centering
    \includegraphics[width=12cm, trim=1.7cm 2.7cm 14cm 0.5cm, clip=True]{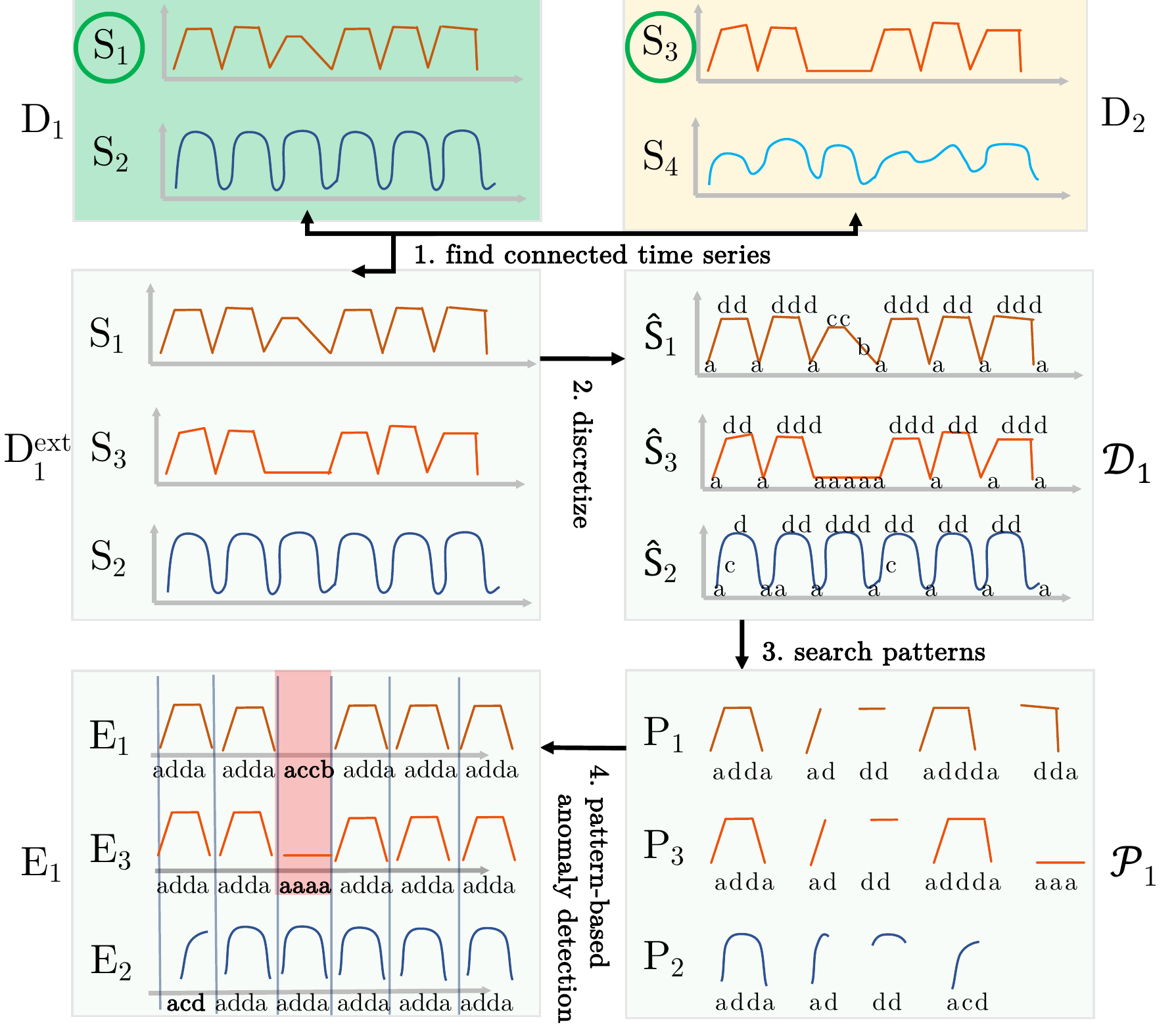}
    \caption{Overview of the proposed \textsc{Bad} method: 1) we search for connected time series, e.g.  sensor $S_1$ of device $D_1$ is similar to $S_3$ of device $D_2$; 2) we extend $D_1$ and include time series $S_3$ from $D_2$ as an additional signal; 3) we transform each time series to a discrete sequence using \textsc{Sax}; 4) we search interesting sequential patterns; 5) we assign an anomaly score based on pattern occurrences, e.g $P_1=adda$ is absent in the 3\textsuperscript{th} window in $E_1$ and $E_3$}
    \label{fig:overview}
\end{figure}

For anomaly detection \emph{efficiency} is important, especially for large networks with many multivariate entities, where the training cost of learning a
separate neural network for each entity is prohibitive and
impedes real-time monitoring. Secondly, \emph{interpretability} is crucial and we want to assist operators in troubleshooting the root-cause and not only to enable them to locate the device and sensor causing
an anomaly during a certain time period. 
For anomaly detection in data series we compute a high anomaly score if \emph{few recurring patterns occur}. \textsc{Fpof}, is one of the first \emph{pattern-based anomaly detection methods} \cite{he2005fp} and recently extended by \textsc{MivPod}~\cite{hemalatha2015minimal} and \textsc{Pbad}~\cite{feremans2019pattern}. All three methods apply a sliding window, discover frequent patterns and compute an outlier or anomaly score based on pattern occurrences in each window. However, there are major differences in: (1) the type of patterns being mined, (2) how to select an interesting pattern, (3) how to compute the anomaly score. 
Existing pattern-based anomaly detection algorithm focus on frequent itemsets \cite{he2005fp}, maximal sequential patterns \cite{feremans2019pattern} or minimal infrequent itemsets \cite{hemalatha2015minimal}. Different to existing methods we mine frequent sequential patterns subject to both \emph{temporal} and \emph{information-theoretical} constraints.  We use a constraint on \emph{minimum length} to avoid  short sequential patterns and on \emph{relative duration}, or cohesion, to ensure that any occurrence of a candidate sequential pattern $X$ contains few gaps \cite{feremans2022}. For instance, both $(a,b,c)$ and $(d,e,f)$ occur in sequence $(a,b,c,x,d,x,x,e,x,f)$, but we only consider the occurrence of pattern $(a,b,c)$. Additionally, this constraint results in a smaller search space and fewer patterns. A second problem with existing pattern-based anomaly detection methods, is that an anomaly prediction is not always explainable. While a  pattern and its occurrences can be inspected if an anomaly is reported using the previously mentioned pattern-based anomaly detection algorithms, it is arguably \emph{not explainable} if there are thousands of patterns. 
Hence, we propose to filter frequent sequential patterns using the \emph{Minimum Description Length (MDL)} principle~\cite{grunwald2007minimum}. 
 That is, we compute how \emph{many bits are saved} by storing the sequence database using  patterns including the description of the patterns themselves \cite{vreeken2019modern}.
 An overview of our context-aware Pattern-\underline{B}ased \underline{A}nomaly \underline{D}etection  (\textsc{Bad}) method is shown in Figure \ref{fig:overview}. 
 

The problem we are tackling is defined as follows: Given a network of multivariate time series (or devices) identify periods of abnormal behaviour.
 Our main contributions are:
\begin{itemize}
 \item A novel pattern-based anomaly detection method. We search for patterns that are frequent and filter interesting patterns based on temporal and information-theoretical constraints. Since, the set of patterns is relatively small and sparse we facilitate human oversight.
    
    \item A novel method to reverse engineering 
 a graph of connected devices. We represent each time series using a fingerprint such that comparing pairwise similarities is both fast and accurate. Thereafter, we prune spurious relations based on patterns on the type of sensor of common occurring relations.
   
    \item  For multivariate time series, 
    we construct a pattern-based embedding for univariate time series and report anomalies by considering the joint probability of patterns occurring in each time series using an isolation forest or \textsc{Fpof} for interpretability~\cite{liu2008isolation,he2005fp}. 
    \item Experimentally we demonstrate the interpretability, efficiency and high anomaly detection performance of the proposed method on large-scale heterogeneous networks. 
\end{itemize}

The remainder of this paper is organised as follows.  In Section~\ref{setting}, we introduce the necessary preliminaries. 
In Section \ref{anomalydetection}, we present a novel pattern-based anomaly detection method. In Section \ref{similar}, we present a method for identifying connected devices based on sensor time series. In Section \ref{together}, we combine both steps for detecting anomalies in multivariate time series. 
In Section~\ref{exp}, we present an experimental evaluation of our method and compare it with state-of-the-art methods. We discuss the related work in Section~\ref{relatedwork} and conclude our work in Section~\ref{conclusion}. 

\section{Preliminaries}
\label{setting}



\subsection{Time series data}
A \emph{time series}, or sensor, is defined as a sequence of measurements $S = (x_1,  x_2, \ldots, x_n)$ where $x_i \in \mathbb{R}$ is sampled at regular time intervals. We denote the type of a sensor using $\mathit{type}(S)$. 

A time series \emph{window}, or interval, is a contiguous subsequence of a time series $S_{i:j}= (x_i,x_{i+1}, \ldots x_j)$ where $1\leq i<j\leq n$. 

Given a window length $l$ and increment $t$ we can transform each time series to a set of \emph{sliding windows}, i.e., $S_{1:l}, S_{1+t:l+t}, \ldots, S_{n-l:n}$ into a continuous sequence database $D$ is, that is
$$D = \{S_{i:j}\ | \ S_{i:j} \in \mathit{sliding\_window}(S,l,t)\}$$
The size of the sequence database is denoted by $|D| = (|S| - l + 1) / t$ where $l$ is the window length and $t$ an increment of the sliding window transform. In practice, we select $t=1$.

An \emph{entity}, or device, consists of $m$ time series, i.e., $D = \{S^1, S^2, \ldots,$ $ S^m\}$ where all time series start and stop at the same time. 
We denote the type of a device using $\mathit{type}(D)$. We assume that each device of the same type has the same amount and types of sensors. 

A multivariate window consists of $m$ subsequences (or dimensions) denoted  $D_{i:j}=\{S^1_{i:j},S^2_{i:j},\ldots,S^m_{i:j}\}$. 

A \emph{network}, $G = (V,E)$ consists of $k$ entities $V(G)=\{D_1, D_2, \ldots, $ $D_k\}$ and a set of edges between entities $E(G)$. We assume that the set of edges is unknown as edges are often dynamic in nature and caused by an unknown underlying data generating process.

\subsection{Sequential pattern mining}    
We use \emph{frequent pattern mining} for discovering recurring behaviour. For pattern mining we requires a database of discrete sequences. Therefore, we transform each time series into a discrete sequence using an extension of Symbolic Aggregate Approximation (\textsc{Sax})~\cite{lin2003symbolic}. 

A \emph{sequence database} $\mathcal{D}$ is created by applying a sliding window on a time series $S$ and transforming each continuous window $S_{i:j}$ to a discrete sequence $\hat{S}_{i:j}$ using \textsc{Sax}, where $$\mathcal{D} = \{\hat{S}_{i:j}\ | \ S_{i:j} \in D: \hat{S}_{i:j} = \textsc{Sax}(S_{i:j},\mathit{paa\_win},\mathit{no\_bins})\}.$$Given a sequence database we can discover frequent unordered or ordered patterns. Many algorithms have been proposed to efficiently mine (non-redundant) frequent patterns 
 and are publicly available in toolkits such as \textsc{Spmf} \cite{zaki2014data,fournier2016spmf}.

A \emph{sequential pattern} $X=(e_1,e_2,\ldots e_p)$ is an ordered list of one or more discretised values or items. 
We use $|X|$ to denote the length of a pattern. 

A sequential pattern  $X$ \emph{occurs} in a discretised sequence, i.e.
$X \prec \hat{S}_{i:j}$, if it is a subsequence of $\hat{S}_{i:j}$ thereby allowing gaps. 
The \emph{cover} of a sequential pattern $X$ in $\mathcal{D}$ is defined as 
$$\mathit{cover}(X,\mathcal{D}) = \{\hat{S}_{i:j}\ | \ \hat{S}_{i:j} \in \mathcal{D} \wedge  X \prec \hat{S}_{i:j}\}.$$
For instance, $X=(a,d,d,a)$ occurs in $\hat{S}=(c,c,\underline{a},\underline{d}, \underline{d}, c, \underline{a})$.

The \emph{support} of a pattern is defined as $\mathit{support}(X,\mathcal{D}) = |\mathit{cover}(X,\mathcal{D})|$. The relative support is defined as  $\mathit{rsupport}(X,\mathcal{D}) = |\mathit{cover}(X,\mathcal{D})|/ |\mathcal{D}|$. A sequential pattern is \emph{frequent} if its support is higher than the user-defined \textit{minimal support} threshold. 

More recently, authors suggest to mine the \emph{top-$k$ most frequent patterns} directly
, which is more intuitive for end-users since choosing \textit{minsup} depends on a database characteristics which are unknown \cite{fournier2013tks,kieu2017mining}. 

An issue with baseline sequential pattern mining algorithms is that they discover many \emph{redundant} and uninteresting patterns. We impose additional \emph{temporal constraints} based on minimal length and the number of gaps to find more interesting patterns in data series \cite{cule2019efficiently}.



A constraint on the \emph{minimal length} defines that for any discovered frequent sequential pattern $X$, the pattern length should be higher than a user-defined threshold, i.e. $|X| > \mathit{min\_len}$ \cite{amphawan2015mining}. By default we set $\mathit{min\_len}=3$ to filter sequential patterns consisting of 1 or two items, since they only cover a small fraction of a sequence and are often frequent due to chance.

A constraint on the \emph{relative duration} determines the cohesion of sequential pattern occurrences \cite{feremans2022}. We impose a constraint on the duration relative to the length of pattern proposed, i.e. $$\mathit{duration}(X,\hat{S}_{i:j})/|X| \geq \mathit{min\_relative\_duration}.$$
For instance, given a sequence $(a,b,c,x,d,x,x,e,x,f)$ the relative duration of sequential pattern $(a,b,c)$ is $\frac{3}{3}$, while for $(d,e,f)$ it is $\frac{6}{3}$. With a relative duration of 1.0 we only count sequential patterns that occur consecutively in each sequence window. By default we set minimal relative duration to 1.2. We remark that by incorporating gap constraints during the mining process, we reduce the search space \cite{masseglia2009efficient, feremans2022}.





\subsection{Symbolic Aggregate Approximation}

\begin{figure}[t!]
    \centering
    \includegraphics[width=7cm,trim=3.5cm 22.5cm 5cm 3cm, clip=True]
    {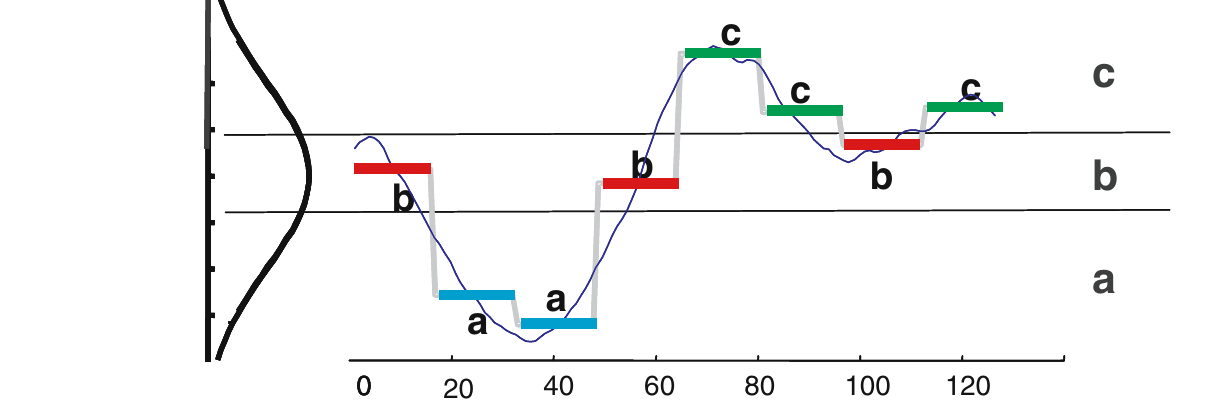}
   \caption{Illustration of the \textsc{Sax} transform where a time series (or time series window) with length $128$ is transformed into the discrete sequence $(b,a,a,b,c,c,b,c)$ of length $8$ using $\mathit{no\_bins}=$ and $\mathit{paa\_win}=16$.}
    \label{fig1}
\end{figure}

We tranform each continuous window $S_{i:j}$ to a discrete sequence $\hat{S}_{i:j}$ using \textsc{Sax} and segment a time series into $n$ pieces, down-sample using averaging and then discretise into $k$ bins.
An example of \textsc{Sax} is shown in Figure \ref{fig1}. We use non-overlapping sliding windows to compute average values. This is controlled using parameter $\mathit{paa\_win}$, which determines how many raw values are averaged, e.g., in the example $\mathit{paa\_win}$ is $16$. Parameter $\mathit{no\_bins}$ controls how many symbols or bins are used for discretisation, e.g., in the example $\mathit{no\_bins}$ is $3$. 

\subsubsection{Binning methods}
For binning several options are possible, i.e., equal-length or equal-width bins or bins selected assuming a normal distribution.  The default binning method used in our experiments is \textit{normal} where we use the same equal-length bins for all windows in one time series, i.e., we compute $\mathit{no\_bins}$ equal-length bins between the lowest and highest value in the entire time series.   
The second strategy is \textit{local}, where we use local bins for each time series window, i.e., we create bins $[a,b,c]$ (or low, medium, high) relative to the minimal and maximal value in one window. This is the strategy originally proposed by \textsc{Sax}~\cite{lin2003symbolic}. As a third alternative, we use \textit{k-means} clustering to bin all time series values into $k$ clusters, where $k$ is chosen to be equal to $\mathit{no\_bins}$.

Each binning approach has its advantages and disadvantages, i.e., using local binning we match the same shape, such as an increase in slope, irregardless of the absolute values. The main difference with the normal strategy is that we compare relative values of symbols and not absolute values. However,  using the normal strategy a certain pattern $X=(a,b,c)$ meaning low, medium, high value (assuming 3 bins) is matched in the entire time series in the same manner and therefore easier to interpret and visualise. Finally, k-means, arguably, leads to a superior binning strategy, compared to equal-length bins. However, we face the problem that clustering is stochastic and can produce different bins in highly similar time series.



\section{Pattern-based anomaly detection}
\label{anomalydetection}


 \subsection{Definition of interesting patterns}
Given a time series $S$ is transformed into a discrete sequence database $\mathcal{D}$ using a sliding window and discretised using \textsc{Sax}. We discover a set of \emph{interesting} sequential patterns $X \in \mathcal{P}$, defines as:
\begin{enumerate}
     \item $|X| > \mathit{min\_len}$
     \item $\mathit{bits\_saved}(X,\mathcal{D})> 0$
     \item $X$ is ranked in the top-$k$ most frequent sequential patterns
     \item The frequency of the pattern $X$ is computed w.r.t. relative duration, i.e. $rank(X) < rank(Y) \leftrightarrow|\mathit{cover_{rdur}}(X,\mathcal{D})| > |\mathit{cover_{rdur}}(Y,\mathcal{D})|$
\end{enumerate}
\subsubsection{Temporal constraints}
The cover of a pattern $X$ w.r.t. relative duration is defined using: \begin{equation*}\begin{aligned}
     \mathit{cover_{rdur}}(X,\mathcal{D}) = 
     \{\hat{S}_{i:j}\ | & \ \hat{S}_{i:j} \in \mathcal{D}: X \prec \hat{S}_{i:j} \ \wedge  \\
     & \frac{\mathit{duration}(X,\hat{S}_{i:j})}{|X|} \ge  \mathit{rdur}\}.
     \end{aligned}
     \end{equation*}
We remark that the maximal length of $X$ is constrained by the \textsc{Sax} representation, i.e. given a sliding window $l$ and $\mathit{paa\_win}$, $|\hat{S}_{i:j}| = l / \mathit{paa\_win}$. 

\subsubsection{Minimal description length}


We are inspired by \emph{MDL} as an additional criterion to individually select sequential patterns. That is, we keep a sequential pattern if its frequent, conforms to temporal constraints, and compresses the data series, i.e., if we need fewer symbols to encode the data series using the pattern (including the pattern itself) than we need to encode the data series literally. The resulting set of patterns $\mathcal{P}$ should describe all regularities in the time series.
Formally, given a sequential pattern $X \in \mathcal{P}$ and its cover $\mathcal{D}_{X} = \mathit{cover}({X,\mathcal{D})}$, we define the number of bits saved using:
\begin{equation*}
    \begin{gathered}
    \mathit{bits\_saved}(X, \mathcal{D}) = \sum_{\hat{S}_{i:j} \in \mathcal{D}_{X}}{\mathit{DL}(\hat{S}_{i:j})}
    - (|X| \cdot log(n) + \sum_{\hat{S}_{i:j} \in \mathcal{D}_{X}}{\mathit{DL}(\hat{S}_{i:j}| \ X)}).
    \end{gathered}
\end{equation*}
Here $\mathit{DL}(\hat{S}_{i:j})$ is the description length to store the covered sequences sequence computed using Huffman coding\footnote{We remark that Huffman coding computes an optimal table of varying-length codes for all sequences in $\mathcal{D}_{X}$. For computing description length we concatenate all covered sequences and ignore the storage cost of  the Huffman table.}. The term $|X| \cdot log(n)$ is the number of bits used to store the pattern, and $\mathit{DL}(\hat{S}_{i:j} | X)$ is computed by replacing the pattern occurrence of $X$ in $\hat{S}_{i:j}$ and then computing the description length of the remaining symbols. 
For example, assuming $X=(a,b,c)$ and $\hat{S}_{i:j}=(x,y,a,b,c,z)$ we compute the description length of $\mathit{DL}(\hat{S}_{i:j}| X) = \mathit{DL}((x,y,*,z))$ where $*$ is an additional symbol to demarcate the location(s) of pattern $X$. In case the occurrence contains gaps we keep gap symbols.  We remark that the number of bits saved is maximal for patterns that are both long and frequent, while frequency is always maximal for short patterns. In preliminary experiments, we found that by filtering patterns where $\mathit{bits\_saved}(X, \mathcal{D}) > 0$, we have an order-of-magnitude less patterns.


\subsection{Efficient algorithm}
\begin{algorithm2e}[h!]
\SetKwInOut{Input}{Input}
\caption{\textsc{Mip}($S$, $l$, $\mathit{paa\_win}$, $no\_bins$, $k$, $\mathit{min\_len}$, $\mathit{rdur}$) Discover sequential patterns that are frequent, satisfy both temporal and MDL constraints}
\Input{Time series $S$, sliding window interval ($l$), \textsc{Sax} parameters ($\mathit{paa\_win}$ and $no\_bins$), number of patterns ($k$), min length pattern (\emph{min\_len}) and min relative duration (\emph{rdur})}
\KwResult{Interesting sequential patterns}
\label{alg:mining}
   \LinesNumbered
    $\mathcal{D} \gets 
    \{\hat{S}_{i:j}| S_{i:j} \in \mathit{sliding\_window(S, l, 1)}: \hat{S}_{i:j} = \textsc{Sax}(S_{i:j},\mathit{paa\_win},\mathit{no\_bins})\}$\\
    $\mathit{stack} \gets \left[\langle \emptyset, \mathcal{D}, \Omega \rangle\right]$\tcp*{Contains current pattern, projection, candidate suffix items} \label{l1}	
    $\mathcal{P}\gets \mathit{make\_heap}(k)$\tcp*{Max $k$ patterns sorted on support}\label{l2}
    $\mathcal{P}_{\mathit{final}} \gets \emptyset $\\
     \While{$\mathit{stack} \neq \emptyset$\label{l3}}{
                $\langle X, \mathcal{D}_{X}, Y \rangle  \gets \mathit{stack}.\mathit{pop}()$\\ \label{l4}
                \If(\tcp*[h]{Stop condition: check if pattern is interesting}){$Y = \emptyset$}{
                
                     \If{$\mathit{support}_{\mathit{rdur}}(X,\mathcal{D}) > \mathit{min\_heap}(\mathcal{P}) \ \mathbf{and} \ |X| \ge \mathit{min\_len}$ \label{l7}}{
                       \If{$\mathit{bits\_saved}(X,\mathcal{D}) > 0$}{
                            $\mathcal{P}_{\mathit{final}} \gets \mathcal{P}_{\mathit{final}} 
 \cup \{X\}$\\
                        }
                        $\mathcal{P}.\mathit{push\_pop}(X)$\\
                    }}
                \Else(\tcp*[h]{Branch and bound: add super-pattern}){
                    \If{$\mathit{support}_{\mathit{rdur}}(X,\mathcal{D}) < \mathit{min\_heap}(\mathcal{P})$}{
                        $\mathbf{continue}$\\
                    }
                    $s_{k+1} \gets Y[0]$\\
                    $X_{k+1} \gets X \oplus s_{k+1}$\\
                    $\textit{stack}.\mathit{push}(\langle X, \mathcal{D}_{X},  Y \setminus \{s_{k+1}\} \rangle)$\\
                    $\mathcal{D}_{X_{k+1}} \gets \textsc{Project}(\mathcal{D}_{X}, X_{k+1}, \mathit{rdur})$\label{l20}\\
                    $Y_{X_{k+1}} \gets \textsc{Candidates}(\mathcal{D}_{X_{k+1}})$\\
                   $\mathit{stack}.\mathit{push}(\langle X_{k+1}, \mathcal{D}_{X_{k+1}}, Y_{X_{k+1}} \rangle)$\label{l21}\\
                }
            }
		$\mathbf{return} \ \mathcal{P}_{\mathit{final}}$
\end{algorithm2e}

We now present and algorithm to mine interesting patterns. The algorithm is an extension of the algorithm proposed by \cite{feremans2022} for mining sequential patterns with respect to temporal constraints. Major differences are that we compute multiple pattern-occurrences for each window add select patterns using MDL.

\textsc{Mine\_Interesting\_Patterns (Mip)}, is shown in Algorithm~\ref{alg:mining}. We start by creating an empty stack. During depth-first recursion we maintain candidate sequential pattern prefixes $X$ and a set of candidate items $Y$. At each iteration we do a divide-and-conquer step, i.e. assume $X=(s_1,\ldots,s_{k})$ and $Y=\{s_a,s_b,s_c\}$ we generate candidate supersequences $X_{k+1}$ by adding every item $s_k \in Y$, i.e. $X_{k+1}=X \oplus s_{k+1} = (s_1,\ldots,s_{k},s_{k+1})$. If no more candidate items exists we evaluate $X$ and check the frequency, minimum length and MDL constraints.  The space of possible varying-length sequential patterns is exponential, thus pruning is extremely important. We limit candidates using three different ways. \begin{itemize}
\item First, if $X$ is infrequent, we know all super-sequences are infrequent and stop early. We use the minimal support in the heap to prune $X$ instead of a fixed value for $\mathit{min\_sup}$ (if fewer than $k$ candidates are in the heap, this is 0). As the number of iterations increase, so does the minimal support of the worst pattern \footnote{Since the minimal support threshold monotonically increases, we trivially guarantee that the final heap contains indeed the top-$k$ most frequent patterns.}. 
\item Second, we use prefix-projected pattern-growth \cite{han2001prefixspan}. Hence, we make use of an additional data-structure, i.e. the \emph{prefix-projection}, which we compute for each candidate during recursion. The projection of the database $\mathcal{D}$ on $X$ contains all sequences covered by $X$ and a pointer to the first and last occurring item $s_k$. In case $X$ occurs multiple times in a sequence, we keep a reference to all positions. We update the prefix-projected database for each new super-sequential pattern, instead of having to scan the entire database, which is more efficient, especially for longer (or low-support) patterns where $|\mathcal{D}_{X}| \ll |\mathcal{D}|$. Additionally, we compute the set of candidates $Y$ to make super-sequential patterns based on the projection. 
\item Third, we further limit the number of candidates based on relative duration. If the minimal relative duration is 1, the set of candidates consists of the next item after $s_k$ in each covered sequence in $\mathcal{D}_{X}$. If the relative duration is higher we have an upper bound of $\lfloor \mathit{max\_len}\times duration \rfloor - \mathit{max\_len}$ on the number of gaps and consequently limit the search to a only a few items after $s_k$ in $\mathcal{D}_{X}$.
\end{itemize}

\subsubsection{Complexity}
The algorithm is memory-friendly, i.e. given the depth-first search strategy of searching for frequent patterns, we maintain at most $\mathit{max\_len}$ candidates. Secondly, we do not actually materialise (or copy) the original sequence database in memory to compute the prefix-projection of the database, but rather compute a \emph{pseudo} prefix-projection thereby storing only indices (or pointers) to covered subsequences and the first and last item position.

The algorithm generates fewer candidates then typically in sequential pattern mining. To enumerate all sequential patterns of size $3$ with $n$ symbols without pruning would generate $n^3$ candidates\footnote{We remark that repeating the same symbol is allowed which is not always the case in sequential pattern mining literature}. Traditional level-wise or column-wise algorithms, would first generate $n$ singletons. Assuming $n'$ singletons are frequent, we would then generate $n' \times n$ candidates of size two. Next, assuming $n''$ frequent patterns of size 2,  we would then generate $n'' \times n$ candidates of length 3. In contrast, we do not consider $n$ super-sequences for each candidate pattern of a certain size, but generate candidates based on items that occur at least once (after the last occurring item of $X$) based on the prefix-projected database.

It is also interesting to compare the complexity with consecutive sequence, motif, or \emph{string mining} algorithms \cite{raza2020accelerating}. To generate all frequent strings of length 3 in a discrete sequence of length $k$ we have only $k-3+1$ candidates, i.e. we would traverse the sequence from left the right using a sliding window of size 3 and an increment of 1. To generate all frequent string of length 3 with at most 1 gap, we have $k-4+1$ windows and need to check $\binom{4}{3}=4$ combinations for each window, or in general, for a pattern of size $l$ with $g$ gaps, we have $(k-(l+g)-1) \times \binom{l+g}{l}$ candidates which is inefficient. We limit the number of candidates in each suffix based on relative duration and leverage the anti-monotonicity of support, thereby combining optimisations from both string and sequential pattern mining.


\subsection{Create an embedding}
Given a set of patterns $\mathcal{P}=\{X_1,X_2\ldots,X_k\}$ we construct an embedding $E_{i:j} \in \mathbb{R}^k$ for each discretised window $\hat{S}_{i:j}$, using
$$E_{i:j}=[f(X_1,\hat{S}_{i:j}), \ldots,  f(X_k,\hat{S}_{i:j})] \quad \text{where}$$
$$f(X_n,\hat{S}_{i:j})= \big \{ 
  \begin{array}{ c l }
    \mathit{rsupport}(X_n,\mathcal{D})  & \quad \textrm{if } X_n \prec \hat{S}_{i:j} \\
    0                 & \quad \textrm{otherwise}
  \end{array},
$$
where we use the \emph{relative support} of a pattern as a feature as opposed to a simpler binary function i.e., 1 if $X_k \prec \hat{S}_{i:j}$ and 0 otherwise\footnote{We remark that a a distance-weighted similarity score is applied by \cite{feremans2019pattern}.}.

A na\"ive algorithm for creating the embedding would search for occurrences for each pattern $X$ in all sequences in $\mathcal{D}$ and has a complexity of $O(|\mathcal{D}| \times |\mathcal{P}|)$. In preliminary experiments we found that the na\"ive algorithm has a longer runtime than pattern mining itself, so its optimisation is also crucial.  
We adopt the prefix-projected data-structure, but for \emph{computing occurrences} of patterns in any data series. For a pattern $X=(s_1,\ldots,s_k)$ we compute the projection for each prefix $(s_1), (s_1,s_2), \ldots$, $(s_1,\ldots,s_k)$ incrementally. The number of sequences matching in each successive projection grows smaller (since support is anti-monotonic), thereby \emph{early abandoning} a sequence when if fails to match a prefix pattern. This method creates the embedding more efficiently by leveraging the \emph{sparsity} of the embedding and temporal constraints. 


\subsection{Compute the anomaly score}

\subsubsection{Isolation forest}
Given the pattern-based embedding for each window $E_{i,j} \in \mathbb{R}^{|\mathcal{P}|}$ we compute anomaly scores using an \emph{isolation forest} \cite{liu2008isolation}. An embedding $E_{i:j}$ has a low anomaly score if many frequent patterns occur and the (average) depth of random trees is relatively high, or conversely a high anomaly score if few frequent patterns occur in the embedding vector for each window. 

\subsubsection{Frequent pattern outlier factor}
To improve interpretability we compute anomalies alternatively using the \emph{frequent pattern outlier factor} (\textsc{Fpof})~\cite{he2005fp}. The anomaly score in for window $\hat{S}_{i:j}$ is computed using 
$$\mathit{fpof}(\hat{S}_{i:j}) = \frac{1}{|\mathcal{P}|} \sum_{X \in \mathcal{P} \wedge X \prec \hat{S}_{i:j}} \mathit{rsupport}(X, \mathcal{D}).$$ 
Here, the outlier factor is high (close to 1) if many frequent patterns (weighted by the relative support) match the current window and low otherwise. For ranking anomaly scores we use $1 -\mathit{fpof}(\hat{S}_{i,j})$.

\section{An efficient algorithm to identify inter-device similarities}
\label{similar}
In the previous section we explained a method to detect anomalies based on interesting patterns. A second goal is to determine edges between different devices based on similar time series. We assume a heterogeneous set of 
$k$ entities $D_1, D_2, \ldots, D_k$ where each entity is of certain type, e.g. a server for computing or storage, an industrial device, a client-oriented device, etc. Moreover, each entity consist of
$m$ sensors or time series, each of a certain type,  e.g. CPU or RAM or disk usage, in going or outgoing network traffic etc.  Our aim is to discover relations, or edges, between device sensors, i.e. two devices having a similar CPU load or network transmission. This is challenging because of following reasons.
\begin{itemize}
    \item Give a large network of devices and sensors, let $N$ denote the total number of time series and $|S|$ the length of time series. It is not scalable to compare each pair of time series, since the runtime is $O(N^2 \cdot |S|)$. 
    \item There is \emph{domain knowledge} related to type , i.e. time series of the same type are often similar to each other, e.g. ``device-A/CPU load''. Vice versa, if only single pair of time series ``CPU temperature'' is similar to ``network bit rate'' this is probably spurious.
    \item Similar to \cite{hyndman2015large} we find that common point-based similarity measures, such as Euclidean distance,  Pearson correlation or Dynamic Time Warping often result in \emph{false positives} in a heterogeneous time series databases. Which is aggravated since time series often exhibit a similar trend, being mostly flat, monotonically increasing, or exhibiting periodical concept drift corresponding to day and night activity.
\item \cite{kinney2014equitability} report that common similarity scores, except mutual information, are not \emph{equitable}.  Since, we have heterogeneous sensor types (with high variety in both distribution and trend) this means that a single threshold on the similarity score, will result in a skew in the degree of connections based on the type of sensors. 

    
\end{itemize}

\subsection{Density-based fingerprint}
First, we generate \emph{density-based fingerprint}. 
We create a grid of $n \times k$ cells for each time series thereby computing the \emph{density}, i.e., the number of samples $x_i$ in each cell. Additionally, we use \emph{logarithmic scaling} to give 
more weight to low-density cells such as \emph{peaks}.  Formally, the distance between two time series $S^i,S^j$ using fingerprinting is defined as
\begin{equation}
\begin{gathered}
\mathit{dist\_fp}(S^i,S^j,n,k) = \\
\sum_{x=1}^{n}\sum_{y=1}^{k} \| \mathit{log}(I_{x,y}(S^i) + 1) - \mathit{log}(I_{x,y}(S^j) + 1) \|_2 
\end{gathered}
\end{equation}
where $I_{x,y}(S)$ denotes the number of sensor values in the cell $(x,y)$. Note that in the limit, i.e., for high values of $n$ and $k$, the fingerprint distance converges towards the Euclidean distance. Because we are dealing with time series, we usually set $n$ using an \emph{interval} and discretise the sensor values into a small number of \emph{bins}. Fingerprints are inspired by \emph{the maximal information coefficient}, which is a similarity metric defined using the mutual information on a discretised grid of values which is \emph{equitable}, i.e. the similarity score is comparable between different types of time series \cite{kinney2014equitability}.

\begin{figure}[b!]
    \centering
    \includegraphics[width=10.5cm, trim=1cm 2cm 1cm 2cm, clip=True ]{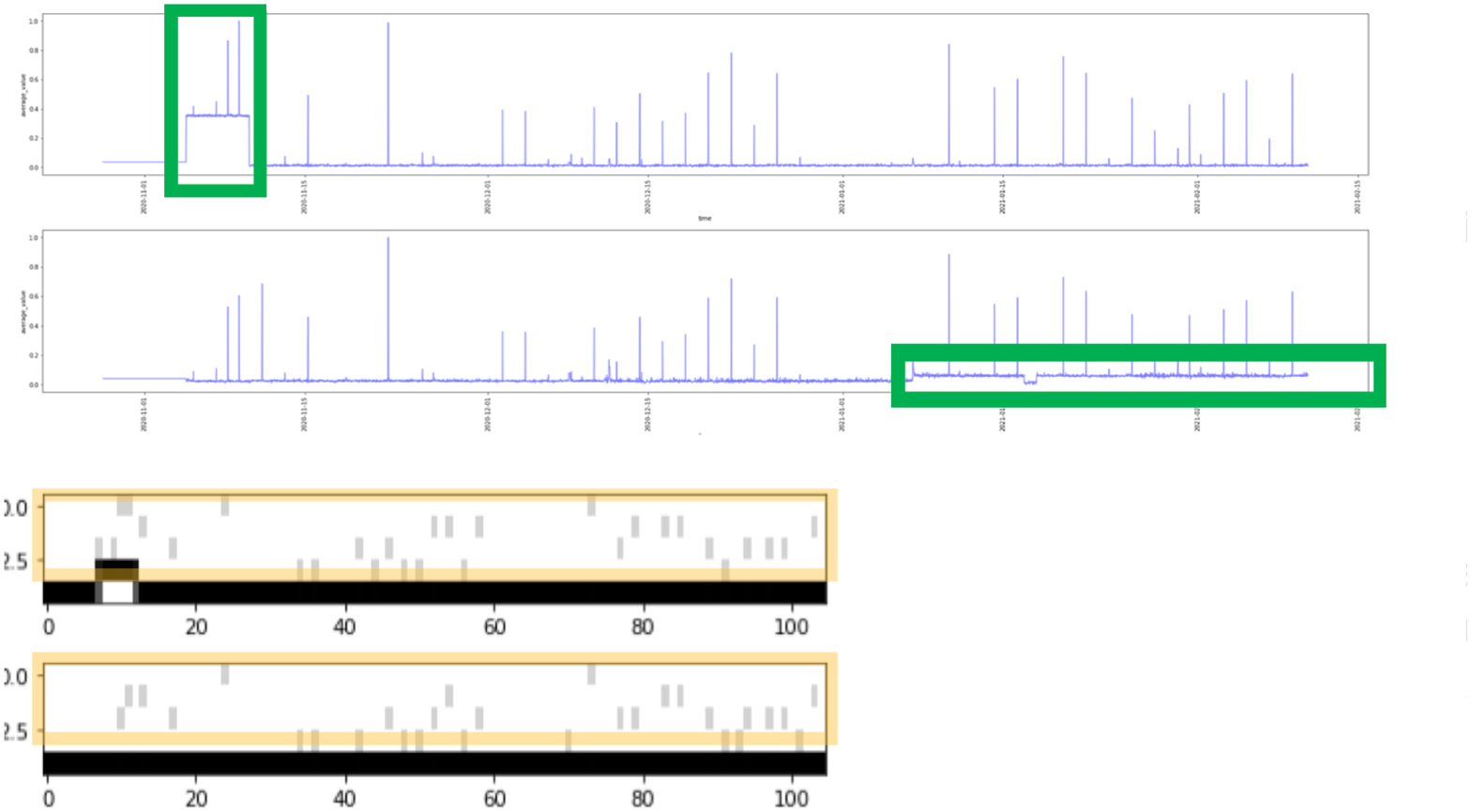}
    \caption{Example of two continuous time series collected over 98 days. Fingerprints are created by counting values in each cell using an interval of $24$ hours and $4$ bins resulting in $103 \times 4$ values instead of $28\ 224$ values}
    \label{fig2}
\end{figure}

An example is shown in Figure \ref{fig2}, where we consider a fingerprint as a low dimensional visual summary for each time series. We highlighted the \emph{concept drift} in green which foils common point-based distance measures, such as Pearson correlation and Dynamic Time Warping. In this example, the  distance between the two fingerprints is low as there is a large correspondence in \emph{peaks}, i.e., light grey cells, and the overall \emph{trend}, i.e., black and white cells.

\subsection{Histogram distances}

The histogram is invariant to phase changes and complementary to distance measures in the time domain. We use histogram distances as a second feature to avoid false positives. 
Given time series $S^i,S^j$ we define the distance between histograms as  
$$\mathit{dist\_hist}(S^i,S^j, l) = \sum_{k=1}^{l} |H^k(S^i) - H^k(S^j)|$$
where $l$ is the number of equal-length bins and $H^k(S)$ denotes the number of sensor values in bin $k$ divided by $|S|$. 

\subsection{Feature-based distance}
We define two time series are similar, i.e., $S^i\sim S^j$, if:
\begin{enumerate}
\item $\mathit{dist\_fp}(S^i,S^j,n,k)  < t_f $
\item $\mathit{dist\_hist}(S^i,S^j,l) < t_h$
\item $(\mathit{type}(S^i),\mathit{type}(S^j)) \in \mathit{frequent\_relations}$
\end{enumerate}
where $t_f$ is a threshold for fingerprint-based distance and $t_h$ is a threshold for histogram-based distance.
First we compute the distance using fingerprinting and histograms between all $\frac{N \times (N-1)}{2}$ pairs to create a symmetric similarity matrix $\mathbf{M} \in \mathbb{R}^{N\times N}$. Next, we search $\mathbf{M}$ to compute the frequency of a relation between two types of sensors as discussed next. 

\begin{algorithm2e}[b!]
\SetKwInOut{Input}{Input}
\caption{\textsc{BuildNetwork}($\mathbf{S}$, $\mathit{interval}$, $\mathit{bins_f}$, $\mathit{t_f}$, $\mathit{bins_h}$, $\mathit{t_h}$, $\mathit{coverage}$) Find connected devices in a network}
\Input{A list of time series filenames $\mathbf{S}$; interval, number of bins and threshold for fingerprint-based and histogram-based distances; coverage for finding frequent relation types}
\KwResult{Set of edges between devices with similar time series}
\label{alg:search}
   \LinesNumbered
   			\tcp{1. Create feature-based representation}
   			\label{line:startprepx}
   			$\mathbf{F} \gets \emptyset$;
   			$\mathbf{H} \gets \emptyset$\\
   			\ForEach{$S^i \in \mathbf{S}$}{
   			    $S^i \gets \mathit{min\_max\_norm}(\mathit{load\_ts}(S^i))$\\
   			    $\mathbf{F}_i \gets \mathit{create\_fingerprint}(S^i, \mathit{interval}, \mathit{bins_f})$\\
   			    $\mathbf{H}_i \gets \mathit{create\_histogram}(S^i, \mathit{bins_h})$\\
   			}
   			\label{line:endprepx}
   			\tcp{2. Compute similarity matrix }
   			$\mathbf{M} \gets 0^{|\mathbf{F}| \times |\mathbf{F}|}$\\
   			\label{line:startsimil}
   			\For{$i \gets 1 \ \KwTo \  |\mathbf{F}|$}{
   				\For{$j \gets i + 1 \ \KwTo \ |\mathbf{F}|$}{
   			        $\mathit{d\_fingerprint} \gets\|\mathbf{F}_i - \mathbf{F}_j\|_2$\\
   			        $\mathit{d\_hist} \gets \|\mathbf{H}_i - \mathbf{H}_j\|_1$\\
   			        \If{$\mathit{d\_fingerprint} < \mathit{t_f} \wedge \mathit{d\_hist} < \mathit{t_h}$}{
   			            $\mathbf{M}_{i,j} \gets \mathit{d\_fingerprint}$
   			        }
   			    }
            } 
            \label{line:endsimil}
            \tcp{3. Find frequent relation types}
            $\mathit{freq\_relations} \gets \textsc{FindFrequentRelationTypes}(\mathbf{M},\mathit{coverage})$\\
            \tcp{4. Create network of connected devices}
            $E(G) \gets \{(D_i, D_j) | \ \exists S^i \in D_i, S^j \in D_j: \qquad \qquad  \qquad \qquad  \qquad \qquad \qquad \qquad \qquad
            \mathbf{M}_{i,j} > 0 \ \wedge  (\mathit{type(S^i)},  \mathit{type(S^j)}) \in \mathit{freq\_relations}\}$\\
            \label{line:endpp}
            $\mathbf{return}\ E(G)$
\end{algorithm2e}

\subsection{Find frequent relation types}
We propose to discover \emph{frequent relation types}. Thereby, we make the assumption that, if the data series between sensors of a certain type are often similar, or frequent, this is a \emph{pattern} in the domain. We remark that the resulting list of frequent relation types also has value for the end-user, i.e. it provides a \emph{comprehensible summary} of the edges in a heterogeneous network.

We represent the similarity matrix $\mathbf{M} \in \mathbb{R}^{N\times N}$ by an \emph{attributed network}, i.e., $G(V,E)$ where each nodes represents a sensor instance labelled by its type and $E(G) = \{(S^k,S^l) \ | \ S^k \sim S^l\}$. Formally, we define the \emph{support} of different types of edges between devices as:
\begin{equation*}
\begin{gathered}
\mathit{support_{RT}}(G, \mathit{type_i}, \mathit{type}_j) = |\{(D_x,D_y) \ | \exists S^k \in D_x, S^l \in D_y:  \\
(S^k, S^l) \in E(G) \wedge \mathit{type(S^k)} = \mathit{type}_i \wedge \mathit{type(S^l)} = \mathit{type}_j\}|.
\end{gathered}
\end{equation*}
The algorithm for discovering frequent relation types is inspired on \emph{separate-and-conquer rule learning} \cite{furnkranz1999separate}. First, we identify the relation type between sensors with the highest support. Next, we remove all devices connected through this relation. In the remaining database, we identify the next most frequent relation type and continue recursively, until we cover, for instance, $90\%$ of time series. 

\subsection{Algorithm}

\begin{algorithm2e}[b!]
\SetKwInOut{Input}{Input}
\caption{\textsc{FindFrequentRelationTypes}($\mathbf{M}$, $\mathit{coverage}$) Find frequent relation types using separate-and-conquer}
\Input{A similarity matrix $\mathbf{M}$ between time series; $\mathit{coverage}$ (\%) parameter, i.e., required percentage of relations in $\mathbf{M}$ to be covered}
\KwResult{Set of frequent relation types}
\label{alg:findfrequent}
   \LinesNumbered
        \tcp{1. Create database}
            $\mathbf{T} \gets \emptyset$\\
             \label{line:startpp}
            \ForEach{$ (i,j) \in \mathbf{M}$}{
                \If{$\mathbf{M}_{i,j} > 0 \wedge \mathit{device}(S^i) \neq \mathit{device}(S^j)$}{
                 $\mathbf{T} \gets \mathbf{T} \cup (\mathit{device}(S^i), \mathit{device}(S^j), \mathit{type}(S^i),  \mathit{type}(S^j))$}
            }
        \tcp{2. Separate-and-conquer}    
            $\mathit{freq\_relations} \gets \emptyset$\\
            $\mathit{min\_cover} \gets |\mathbf{T}| * (1 - \mathit{coverage})$\\
            \While{$|\mathbf{T}| > \mathit{min\_cover}$}{
                $\mathit{freq\_rel} \gets \underset{(t_1,t_2) \in \pi_{\mathit{type}_1,\mathit{type}_2}(\mathbf{T})} {\mathit{argmax}} \mathit{support_{RT}}(T, t_1,t_2)$\\
                \label{line:toprel}
                $\mathit{freq\_relations} \gets \mathit{freq\_relations} \cup \mathit{freq\_rel}$\\ 
                $\mathit{device\_pairs} \gets \pi_{dev_1,dev_2} ( \sigma_{\mathit{type}_1 = t_1 \wedge \mathit{type}_2 = t_2}(\mathbf{T}))$\\
                $\mathbf{T} \gets \mathbf{T} \ \setminus \ \{ r \ | \  r \in \mathbf{T}: (r.dev1,r.dev2) \in \mathit{device\_pairs}\}$
            }
            $\mathbf{return} \text{ } \mathit{freq\_relations}$\
\end{algorithm2e}

The main \textsc{BuildNetwork} algorithm is shown in Algorithm~\ref{alg:search}. We load time series one by one, pre-process them and compute a memory-friendly representation consisting of one fingerprint and histogram for each time series~(line~\ref{line:startprepx}-\ref{line:endprepx}). Next, we compute distances between all $N^2/2$ candidates and create a similarity matrix $\mathbf{M}$ (line~\ref{line:startsimil}-\ref{line:endsimil}).


In Algorithm~\ref{alg:findfrequent} we show the algorithm for discovering frequent relation types based on \emph{coverage}. First, we join all pairs of discovered similar time series with metadata, i.e., the device and type of both time series. Next, we identify the relation type between time series with the highest support (line \ref{line:toprel}). Next, we identify all pairs of devices connected through this frequent relation type and remove all edges between these devices. We keep on mining the next most frequent relation types recursively until we cover the required number of the discovered edges between time series as determined by the $\mathit{coverage}$ hyper-parameter. We remark that a pair of devices is often connected based on multiple correlated time series, however by removing pairs of devices at each iteration we focus on \emph{non-redundant} relation types by computing the support of a relation type after filtering the device pairs connected through higher support relation types. 

Finally, we \emph{build the network} of connected devices based on the similarity matrix and after filtering on the discovered frequent relations as shown in Algorithm \ref{alg:search} (line \ref{line:endpp}).  

\subsubsection{Complexity}
The main bottleneck is the computation of the distance function between all $N^2/2$ candidates to create a symmetric similarity matrix $\mathbf{M} \in \mathbb{R}^{N\times N}$. The proposed feature-based similarity representation has a linear space complexity of $O(n \times k + l)$. A grid of $n \times k$ values for the fingerprint (as determined by the hyper-parameters $interval$ and $bins_f$), and $l$ bins for the histogram (as determined by $bins_h$). Clearly, keeping $n\times k + l \ll |S|$ values in memory is efficient. The algorithm for discovering frequent relation types requires $O(N^2)$ steps in in the worst case, however, in practice the runtime is only a few seconds.

\section{Pattern-based anomaly detection in a network of multivariate time series}
\label{together}
Given the 
methods described in Section \ref{anomalydetection} and \ref{similar}, we are now able to tackle the main problem of detecting context-aware anomalies at the entity level.  The main \textsc{Bad} algorithm is shown in Algorithm \ref{alg:ipbad}. We characterise anomaly detection based on the following three tasks where we compute an anomaly score for: 
\begin{enumerate}
    \item a window $S_{i:j}$ in a \emph{univariate time series}
    \item a window $D_{i:j}$ in a \emph{multivariate time series} or \emph{entity}
    \item a window $D_{i:j}$ in an entity with regards to the \emph{network context}. 
\end{enumerate}   
For instance, for the last case we could discover an edge if the transmitting network bit rate $S^k$ of device $D_p$ is highly correlated to the receiving network bit rate $S^l$ of device $D_q$ and report an anomaly at $D^q_{i:j}$ (or $D^k_{i:j}$) if this is not the case.

\begin{algorithm2e}[t!]
\SetKwInOut{Input}{Input}
\caption{\textsc{Bad}($V$, $D_q$, $\Omega_{network}$, $\Omega_{discrete}$, $\Omega_{mining}$, $\mathit{use\_fpof}$) \text{Efficient} method to discover anomalies in a multivariate network}
\Input{A set of entities $V$ where each entity $D$ has $m$ time series (or sensors); $D_q$ a target entity for anomaly detection; hyper-parameters for each component; method to compute anomalies, i.e. \textsc{Fpof} or isolation forest} 
\KwResult{Anomaly score for each multivariate window $D_{i:j} \in D_q$, or for each univariate window $S^k_{i:j}$ in each sensor $S^k \in D^q$}
\BlankLine
\label{alg:ipbad}
   \LinesNumbered
        \tcp{1. Discover relations between devices}
        $\mathcal{S} \gets \{ S \ | \ \exists \ D \in V: S \in D \} $ \\
        $E(G) \gets \textsc{BuildNetwork}(\mathcal{S}, \Omega_{network})$\\
        $D^q_{\mathit{ext}}= D_q \cup \{S^j \ | \  \exists \ S^i \in D_q, \ S^j \in D_k:  (D_q,D_k) \in E(G) \wedge  S^i \sim S^j\}$\\
        \tcp{2. Create a pattern-based embedding}
        \For{$\text{sensor}\ S^k \in D^q_{\mathit{ext}}$}{
            $\mathcal{P}^k \gets \textsc{Mip}(S,\Omega_{discrete},\Omega_{mining})$ \tcp*{2.1 Mine interesting patterns}
            $E^k \gets \textsc{CreateEmbedding}(\mathcal{P},S^k)$ \tcp*{2.2 Create embedding per sensor}
        }
        \tcp{2.3 Compute anomalies}
        \If{$\mathbf{not} \ \mathit{use\_fpof}$}{
            $\mathbf{E}^{1,\ldots,m} \gets \emptyset$\\
            \For{$\text{window} \ (i,j) \in D^q_{\mathit{ext}}$}{
                $\mathbf{E}^{1,\ldots,m}_{i:j} = \mathit{concatenate}(\{E^k_{i:j} \ | \ S^k \in D^q_{\mathit{ext}}\})$
            }
            $\mathbf{return}\ \textsc{IsolationForest}(\mathbf{E})$
        }
        \Else{
           $A \gets \text{init matrix of anomaly scores for each sensor and window}$\\
           \ForEach{$\text{sensor} \ S^k \in D^q_{\mathit{ext}}$}{
            $A[k] \gets \textsc{Fpof}(\mathcal{P}^k,E^k)$
           }
            $\mathbf{return}\ A$\\
        }	   
\end{algorithm2e}

\subsection{Context-aware entity anomaly detection}
First, we identify connected devices using feature-based similarity as explained in Section \ref{similar}. After we have identified connected devices we classify anomalies in a  \emph{context-aware} manner.

For each device $D$, we create a new collection $D_{\mathit{ext}}$ where we add all time series of the device and similar time series of connected devices, that is 
\begin{equation*}
    \begin{aligned}
    D_{\mathit{ext}}= D \cup \{S^j \ | \  \exists \ S^i \in D, \ S^j \in D_k:  (D,D_k) \in E(G) \wedge  S^i \sim S^j\}.
    \end{aligned}
\end{equation*}
Next, we mine patterns and create an embedding for each time series  and concatenate the embedding vectors. For instance, if $S^i$ and $S^j$ are similar (e.g., the sending and receiving bit rates are similar) we expect that frequent patterns of $\mathcal{P}^i$ and $\mathcal{P}^j$ co-occur causing a high anomaly score if this is not the case. 

\subsection{Anomaly detection in multivariate time series}
For entity-level anomaly detection we discover patterns in each time series \emph{independently} as time series are of different types and an entity $D_{\mathit{ext}}$ (or $D$) is in essence a \emph{heterogeneous} collection. 

\subsubsection{Isolation forest}
Given an entity $D_{\mathit{ext}}=\{S^1,\ldots,S^m\}$ we discover patterns in each time series separately and concatenate the discovered embedding vectors in each time series $S^k$ for each window $(i,j)$ using
$$E^{1,\ldots,m}_{i:j} = \mathit{concatenate}(\{E^k_{i:j} \ | \ S^k \in D\}).$$
Next, we use an isolation forest thereby predicting a window $D_{i:j}$ as anomalous if one or more sensors behave unexpectedly, i.e., if the joint occurrence of patterns is unusual. The advantage of an isolation forest is that dependencies between patterns are captured, i.e., if $X_1 \in \mathcal{P}^1$ and $X_2 \in \mathcal{P}^2$ always occur together, then an anomaly would be flagged if only $X_1$ occurs, which is especially important when combining patterns in multivariate time series. 


\subsubsection{Frequent pattern outlier factor}
Alternatively, we compute the anomaly score \emph{independently} for each time series in $D_{\mathit{ext}}$. Ignoring interactions between patterns in different time series can result in a lower accuracy. However, the anomaly score can be inspected on the level of the individual time series (or connected time series). For instance, if at window $D_{i:j}$ only time series $S^k,S^n$ and $S^l$ have a high anomaly score while others are normal this \emph{localisation} information is useful to the end-user~\cite{su2019robust}.

\section{Experiments}
\label{exp}

In this section, we begin by evaluating our method for identifying inter-device similarities, before comparing our anomaly detection method to state-of-the-art algorithms on several benchmark datasets. Our implementation is open-source and publicly available\footnote{\url{https://bitbucket.org/len_feremans/pbad\_network/}}. We also provided an online tutorial notebook in Python with many examples \footnote{\url{https://lfereman.github.io/}}.

In Table \ref{table:dataset} we show more details on the broadcasting video network dataset. This network consists of 315 entities and $8\,805$ time series. Each sensor was collected over 98 days where the mean value is stored every 5 minutes resulting in $28\,224$ values, i.e. telemetry and physical sensor data association with a single device such as the CPU usage, CPU temperature, memory consumption, disk usage and input and output network bitrate. Each of the time series belong to one of 5 types of entities and there are in total 49 distinct types of time series.  For instance, there are 134 entities of type A where each entity of type A has at most 11 sensors of a certain type.
 
For pre-processing we cap outlier values based on the 99\% quantile and apply min-max normalisation. Many time series consist of straight lines and are less of interest. Hence, we detect straight lines by fitting a linear function using least squares to the time series and remove time series if the Pearson correlation coefficient is higher 0.98.

\subsection{Telecommunications network dataset}

\subsection{Parameter selection for the discrete representation}
The accuracy of any pattern-based anomaly detection technique is highly dependent on reasonable values for parameters determining the discrete representation. In this section we discuss how to select parameters. 

\begin{table*}[h!]
\centering
\begin{tabularx}{0.4\textwidth}{ccc}
\toprule
\textbf{Entity type} & \textbf{\# entities} & \textbf{\# time series}\\
 \hline
 
A & 134 & 11\\
C & 60 & 20\\
S & 73 & 12\\
T & 27 & 2\\
D & 21 & 4\\
\bottomrule
\end{tabularx}
\caption{The details of the data extracted from a broadcasting video network. The network consists of $315$ entities of $5$ types and $8\,805$ time series, or sensors, of $49$ types. Each time series was collected over 100 days.}
\label{table:dataset}
\end{table*}

\subsubsection{Sliding window parameters}
We suggest to always use an \textit{increment} of 1. An increment of 1 minimising the risk that patterns do no match a window because of arbitrary boundaries induced by semi-overlapping sliding windows. 
The \textit{interval} $l$ depends on the domain but the order of magnitude of the interval, i.e., days, hours or seconds depends on the characteristics of the dataset such as the \emph{sampling rate} and the total length of the time series. In the telecom datasets we have $12$ values logged every hour, hence an interval in hours is reasonable, i.e., 2, 8, 24 or 48 hours. 
We remark for setting the interval, a measure of the granularity in time, there is a \textit{trade-off}, i.e., if the interval is selected too course-grained anomaly detection is high, but meaningless since anomaly scores cover a large period. Vice versa, if it is too fine-grained we have the risk of too few data points and low accuracy. 

\subsubsection{\textsc{Sax} parameters}
For \textsc{Sax} we have two parameters that control how many raw values are averaged ($\mathit{paa\_win}$) before binning, and how many bins or the alphabet size ($\mathit{no\_bins}$) are used. The length of the discrete sequence is given by $$|\hat{S}_{i:j}| = \mathit{no\_symbols}=|S_{i:j}| / \mathit{paa\_win} = l/\mathit{paa\_win}.$$
We suggest to vary $\mathit{no\_bins}$ and $\mathit{no\_symbols}$ between 5 and 20 and propose defaults values of $\mathit{no\_bins}=5$ and $\mathit{no\_symbols}=10$. The rationale behind this is that long sequential patterns, i.e. consisting of more than 20 symbols are typically infrequent. Additionally, we want a good coverage of patterns, i.e., a frequent pattern of length 5 will cover $50\%$ of the symbols in a discrete sequence of length 10. If labels are available, we suggest to use grid search.
If labels are not available, we suggest to plot both the time series, time series after discretisation and the pattern-based embedding and use our human visual expertise to select reasonable settings. 
We have a \textit{trade-off}, i.e., if $\mathit{no\_bins}$ is high and $\mathit{paa\_win}$ is small the discrete representation will have high \textit{fidelity} to the original time series, but few frequent patterns, i.e., most subsequences will occur infrequently, especially if there is more noise. Vice versa, if too few bins are selected and a high value for $\mathit{paa\_win}$ we risk that the discrete sequences are short and no longer resembling the original time series. 

\subsection{Efficiency and accuracy of \textsc{BuildNetwork}}

\subsubsection{Parameter Selection}
Since there is no ground truth we select parameters based on expert knowledge and visual confirmation. For fingerprinting, we set the interval to 24 hours and use 5 bins for discretisation. For the histogram creation, we use 100 equal-length bins. We set the threshold for fingerprinting using the 99\% quantile and for histogram distances using the 95\% quantile. The coverage parameter for filtering on frequent relations is set to 95\%. 

\subsubsection{Results for frequent relation types} 
We find $9\,458$ raw pairs of similar time series using these parameter settings. By filtering on intra-device similarities, edges based on known groups of devices, and frequent relation types, only $351$ edges remain. 
In Table \ref{tab:frequenttypes} we show the top-6 of the total $28$ frequent relation types which is much smaller than the $\binom{49}{2} = 1176$ theoretically possible combinations. These frequent relations are a good starting point to \emph{comprehend} this complex network of heterogeneous devices and sensors. 

\begin{table}[b!]
\centering
\begin{tabularx}{0.7\textwidth}{p{8cm} p{2cm}}
\toprule
\textbf{Type} & \textbf{\# edges}\\
\toprule
A/System load & 91\\
A/CPU load & 52\\
A/IP bit rate & 36\\
A/TS rate & 29\\
A/User load & 24\\
A/CPU System - A/CPU system time load & 10\\
\bottomrule
\end{tabularx}
\caption{Most frequent relation types between devices
}
\label{tab:frequenttypes}
\end{table}

\subsubsection{Runtime performance} 
Computing the similarity of all $78\cdot10^6$ possible pairs \emph{takes just $8.5$} minutes or $59.0$ ms per instance to search in $8\,708$ time series after excluding straight lines. 


\subsubsection{Comparing the accuracy of similarity measures} 
We compare the proposed fingerprint-histogram distances against \textsc{Paa} \cite{keogh2001dimensionality} and the Pearson correlation coefficient after sampling each time series to 2000 values. 
 To evaluate performance we compute precision, recall and F1. Since we do not have labels of similar time series, we pick 50 time series, search for the top-5 most similar time series for each of them, and manually assign labels based on domain knowledge. The results are shown in Table \ref{tab:comparesim}. We find that fingerprinting combined with histogram distances has a \emph{higher precision} and \emph{F1} but \emph{lower recall} compared to \textsc{Paa} and that the performance of Pearson correlation is relatively low. 
 However, we argue that for context-aware anomaly detection high precision is of more importance than high recall.  

\begin{table}[b!]
\centering
\begin{tabularx}{0.7\textwidth}{lrrr}
\toprule
\textbf{Method} & \textbf{Precision} & \textbf{Recall} & \textbf{F1}\\
\midrule
\makecell{Fingerprinting and histogram} & $\mathbf{0.971}$ & $0.779$ & $\mathbf{0.864}$\\ 
\makecell{Piecewise aggregate approximation} & $0.706$ & $\mathbf{0.895}$ & $0.789$\\
\makecell{Pearson correlation coefficient} & $0.462$ & $0.639$ & $0.536$\\
\bottomrule
\end{tabularx}%
\caption{Accuracy of time series similarity identification in the broadcasting video network dataset}
\label{tab:comparesim}
 \end{table}

\subsubsection{Qualitative performance}
\begin{figure}[t!]
    \centering
    \includegraphics[width=12.5cm, trim=1.8cm 0.8cm 1.7cm 0.8cm, clip=True]{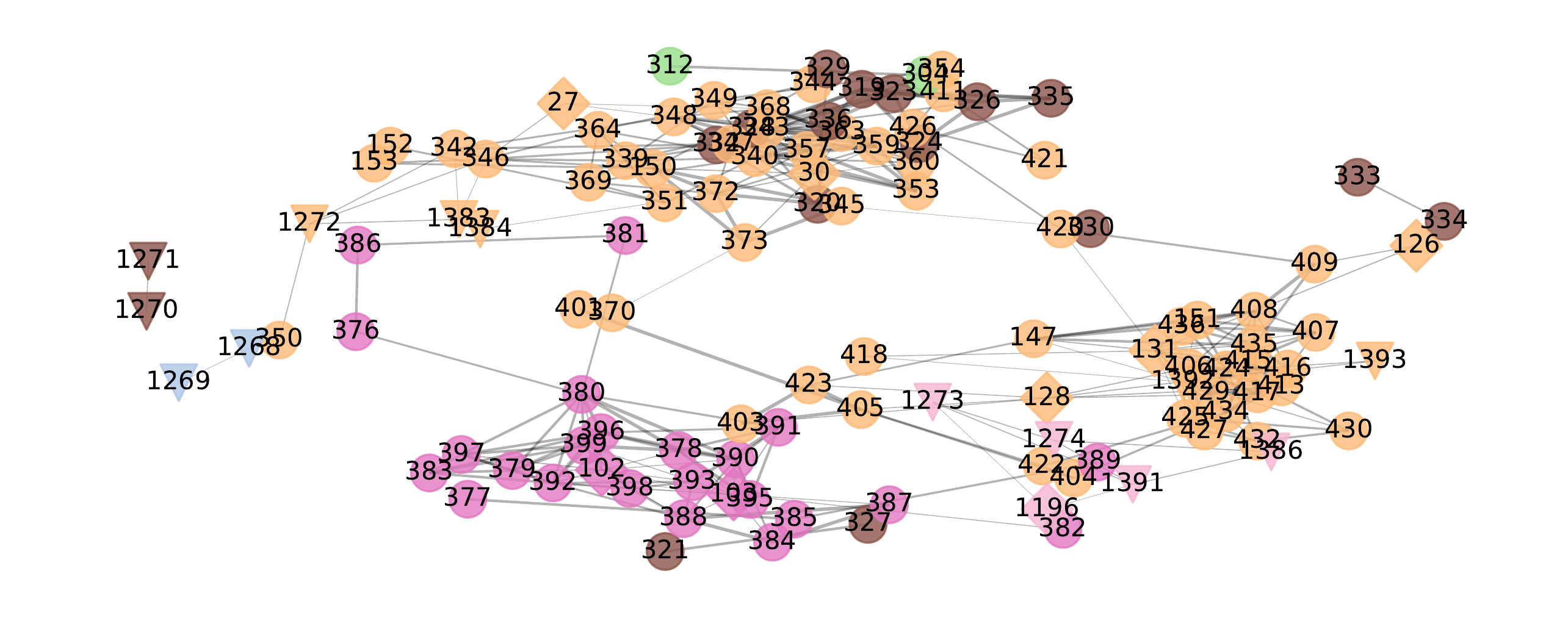}
    \caption{Discovered network of broadcasting video devices. The colour of each node represents a known cluster and the shape the type of device}
    \label{fig:network1}
\end{figure}

In Figure~\ref{fig:network1} we show the resulting network visualised using the NetworkX  package. We find 4 large dense clusters of mostly ``type A'' devices which are a type of card located inside the same ``type C'', so, unsurprisingly, they have multiple similar sensors and a connection with a few ``type C'' elements. We also discover interesting smaller clusters, e.g., ``type D'', and ``type C'' devices are often connected using bit rate. Furthermore, we observe that most discovered relations are between devices where the identifier is near, e.g., device 1268 is connected with 1269. This validates that the proposed algorithm for identifying edges using only time series data matches with the prior knowledge that devices with a near human-assigned identifier are often related.
 
\subsection{Efficiency and accuracy of pattern-based anomaly detection on benchmark datasets}
Since we have no labels for the network dataset, we evaluate our anomaly detection method on univariate and multivariate benchmark datasets. 

\subsubsection{Baselines}
We compare \textsc{Bad} with two state-of-the-art pattern-based anomaly detection methods: \emph{Frequent pattern based outlier factor} (\textsc{Fpof}) computes an outlier factor based on the number of frequent closed itemsets~\cite{he2005fp}, and \emph{Pattern-based anomaly detection} (\textsc{Pbad}) uses an isolation forest trained on the embedding of raw sensor values, closed frequent itemsets and sequential patterns~\cite{feremans2019pattern}. Additionally, we compare with \emph{Isolation forest} that uses a forest trained on the original sensor values in each window~\cite{liu2008isolation} and \textsc{Bad-Fpof} which uses the frequent pattern based outlier factor instead of an isolation forest to facilitate interpretability. 


\subsubsection{Evaluation protocol}
Given a time series $S$ with a number of labelled timestamps, we divide the time series into fixed-sized sliding windows and compute an anomaly score for each window. Similar to the evaluation protocol of  \textsc{OmnyAnomaly}~\cite{su2019robust}, we adopt the \emph{point-adjust} approach where we consider an anomaly detected if it occurs within the boundary of a sliding window. Each method requires the setting of a threshold and we assume an Oracle that reports the best possible F1 value by varying the threshold. Since the isolation forest is not deterministic we run each method 5 times using the same parameters and report the mean values. 

\begin{table*}[h!]
\centering
\begin{tabularx}{0.7\textwidth}{l *{6}{l}}
\toprule
\textbf{Dataset} & \textsc{IsoForest} & \textsc{Fpof} & \textsc{Pbad} & \textsc{Bad} & \textsc{Bad-Fpof}\\
\midrule
\texttt{Temp} & 0.927 & 0.875 & 0.776 & \textbf{0.948} & 0.881\\
\texttt{Taxi} & 0.782 & 0.718 & 0.807 & \textbf{0.851} & 0.630\\
\texttt{Latency} & 0.827 & 0.653 & 0.864 & \textbf{0.901} & 0.845\\
\texttt{Water 1} & 0.686 & 0.333 & 0.755 & \textbf{0.787} & 0.549\\
\texttt{Water 2} & 0.528 & 0.450 & \textbf{0.596} & 0.539 & 0.452  \\
\texttt{Water 3} & \textbf{0.791} & 0.321 &  0.615 & 0.567 & 0.447\\
\texttt{Water 4} & 0.561 & 0.661 & \textbf{0.760} & 0.666 & 0.663\\
\emph{Avg. rank univariate} &  3 & 4.4 & 2.1 & \textbf{1.6} & 3.8\\
\midrule
\texttt{SMD} & 0.949 & n/a & n/a & \textbf{0.966} & 0.957\\
\texttt{SMAP} & 0.708 & n/a & n/a & \textbf{0.775} & 0.733\\
\emph{Avg. rank multivariate} &  3 &  n/a &  n/a & \textbf{1} & 2\\
\bottomrule
\end{tabularx}
\caption{F1 score of various methods on $7$ univariate datasets and $2$ multivariate datasets consisting of $28$ and $54$ entities respectively, 
where we report the mean result
}
\label{table:results_univariate}
\end{table*}

\subsubsection{Parameter selection}
Each method has the same preprocessing steps which include setting the window size $l$ and increment $t$ to create sliding windows and parameters for transforming to a symbolic representation such as the \emph{paa\_window} for reducing the dimensionality and the number of equal-length \emph{bins} which we optimise using \emph{grid search}. For the \textsc{IsolationForest}, we use 500 \emph{trees}. For \textsc{Pbad} we set $\mathit{min\_len}$ to 2 and the \emph{minimal support} to $0.01$. For \textsc{Fpof}, we set  the \emph{minimal support} to $0.01$. For \textsc{Bad}, we set the number of patterns $k$ to $10\,000$, $\mathit{min\_len}$ to 2 and the \emph{relative duration} to $1.2$.

\subsubsection{Datasets}
For the univariate test case, we use 7 datasets. 
\texttt{Temp}, \texttt{Latency} and \texttt{Taxi} are smaller datasets from the Numenta Anomaly Benchmark~\cite{lavin2015evaluating}.   
\texttt{Water} 1 to 4  is a proprietary dataset that contains the average consumption of water in 4 branches of a retail company where the goal is to detect abnormal water consumption~\cite{feremans2019pattern}. The water consumption is logged every 5 minutes for two years resulting in $170\,000$ values. For the multivariate test case, we use the Server Machine Dataset (\texttt{SMD}) which consists of 28 entities, 
with 38 sensors each, containing $56\,960$ values~\cite{su2019robust}. The semantics are similar to our broadcasting video network with 
data such as CPU load, RAM usage, etc. Soil Moisture Active Passive (\texttt{SMAP)} is a dataset from NASA and consists of 54 entities each monitored by 25 metrics \cite{hundman2018detecting}.

\subsubsection{Compare the accuracy with baselines}
Table~\ref{table:results_univariate} shows the F1 obtained by each method on each dataset. Results in bold highlight the best performing method on each dataset. We find that \textsc{Bad} has the lowest average rank on univariate datasets thus outperforming existing state-of-the-art methods such as \textsc{IsoForest}, \textsc{Pbad} and \textsc{Fpof}. We find that both \textsc{Bad} and \textsc{Bad-Fpof} perform better than \textsc{IsoForest} on the multivariate datasets. We report no results for \textsc{Fpof} and \textsc{Pbad} on the multivariate datasets as the run times are prohibitively long.  

We compare with the reported results of recent neural-network based methods, such as \textsc{OmniAnomaly} and \textsc{Usad}~\cite{audibert2020usad}. On the \texttt{SMD} dataset, we find the average F1 value of \textsc{Bad} is $0.966$ $(\pm 0.002)$ which is higher than 
\textsc{OmniAnomaly} (0.962) and \textsc{Usad} (0.946). However, the F1 value for \texttt{SMAP} is lower, i.e. $0.775$ for \textsc{Bad}  versus $0.853$ for \textsc{OmnyAnomaly} and $0.863$ for \textsc{Usad}. The relatively worse performance on \texttt{SMAP} is possibly explained by the special nature of this dataset, where time series often consist of only peaks alternating between two values. We remark that all existing methods ignore context when considering anomaly detection in a network of devices. 

\subsubsection{Runtime performance}
Concerning runtime performance we find that \textsc{Pbad} is considerably slower and requires more than 1.5 hours to complete on each \texttt{Water} dataset. The minimum support threshold results in an unpredictable and larger set of patterns.
In \textsc{Bad} using default parameters 
results in fewer patterns (800 instead of 5000 for the \texttt{Water} dataset) and both pattern discovery and the creation of the embedding are more efficient resulting in a total runtime of less than a few minutes. On \texttt{SMD} 
the total time for pre-processing, learning the model and making predictions was just 37 \emph{minutes} (on a laptop).  
This is in stark contrast to \textsc{OmniAnomaly}, that required more than two hours to train a model for a \emph{single} device on the \texttt{SMD} datasets. 
We conclude that \textsc{Bad} and \textsc{IsoForest} scale to long time series and collections thereof.

\subsubsection{Qualitative evaluation}
\begin{figure*}[h!]
    \centering
    \includegraphics[width=16cm, trim=0cm 4cm 0cm 4cm, clip=True ]{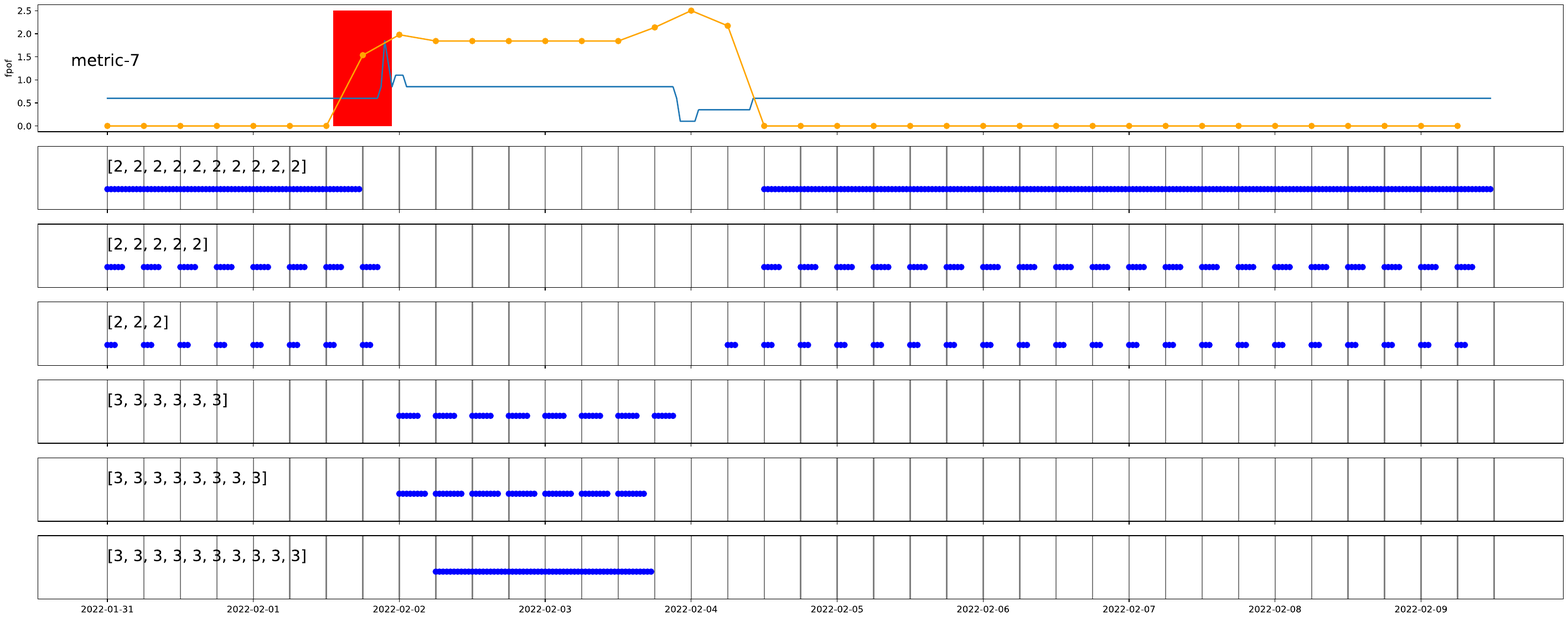}
    \caption{Example of a discrete representation of metric-7 from the \texttt{SMD} dataset where we show the pattern-based embedding. The pattern mining algorithm 
    discovers only 6 patterns. We show the occurrences of each pattern and remark that in this case a high anomaly score (red) is correlated with an absence of pattern occurrences.}
    \label{fig:example2}
\end{figure*}

For decision support, we use \textsc{Bad-Fpof} to allow \emph{localisation} by computing the \textsc{Fpof}-based anomaly score based on patterns discovered in each time series. In Figure \ref{fig:example2} we visualise the pattern-based embedding of the 7\textsuperscript{th} sensor of the first device in the \texttt{SMD} dataset where the \textsc{Fpof} score is based on only 6 patterns. 
The known anomaly is located after the first series of $X_1=[2,2,2,\ldots]$ occurrences and before occurrences of $X_6=[3,3,3,\ldots]$, which is also when few patterns occur causing a high anomaly score. 


\begin{figure*}[h!]
    \centering
    \includegraphics[width=16cm, trim=0cm 3.5cm 0cm 3.5cm, clip=True ]{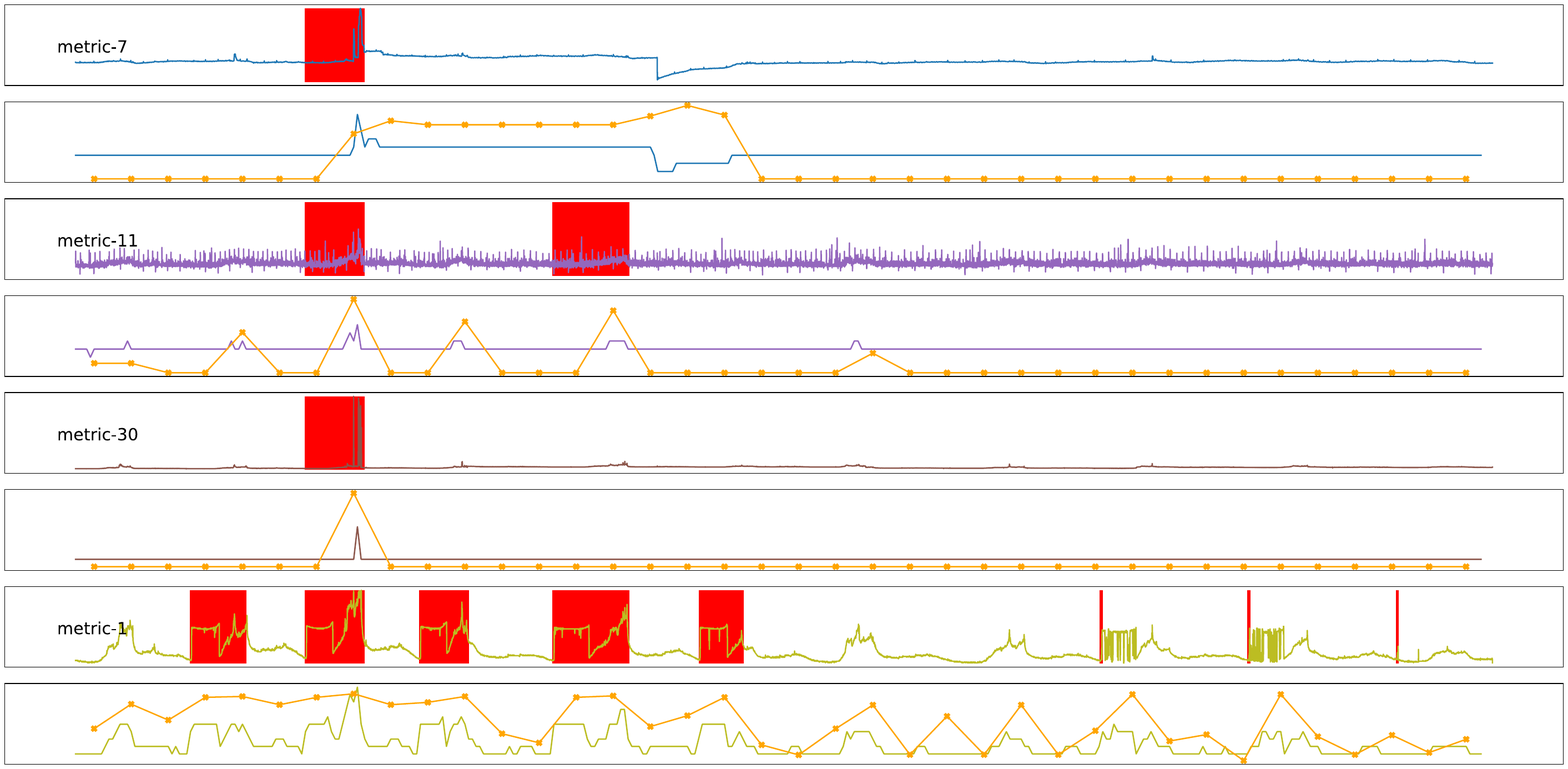}
    \caption{Example of localised anomaly scores of 4 time series from a single device in the \texttt{SMD} dataset using \textsc{Bad-Fpof} . We show the time series before and after discretisation, the known sensor-specific anomalies (in red), and the predicted anomalies (in orange)}
    \label{fig:example1}
\end{figure*}

In Figure~\ref{fig:example1} we show 4 time series from the first device in the \texttt{SMD} dataset. For \texttt{SMD} we have labelled anomalies specific to a subset of time series highlighted in red. Below each time series we show the discrete representation computed using an interval of 6 hours, 10 symbols and 10 equal-length bins and the \textsc{Fpof} score in orange. We remark that there is high correspondence between known and predicted anomalies.

\section{Related work}
\label{relatedwork}

We have shown that the proposed similarity measure that considers the trend, peaks and the distribution of values results in a 
simple efficient method that produces fewer false positive than \textsc{Paa}. For comparing time series more advanced techniques have been proposed, however, few scale to large time series databases. Finally, the approach for filtering spurious relations based on the type of sensor is of specific interest to intra-device similarities and has not been considered before.

For pattern-based anomaly detection we have shown that \textsc{Bad} is more accurate and efficient compared with state-of-the-art methods, such as \textsc{Fpof} and \textsc{Pbad}. Related work, such as \textsc{MivPod} \cite{hemalatha2015minimal} was shown to be less accurate than \textsc{Pbad}. 

In contrast to prior work in sequential pattern mining, we are interested in patterns that are both frequent, cohesive and compress the data series, while previous work considers only one or two aspects, i.e. \cite{vreeken2011krimp, lam2014mining} and \cite{shokoohi2015discovery} discovers a set of patterns the compresses the dataset best, while \cite{feremans2022} discover frequent and cohesive patterns for data series. 

Recently, authors proposed to make complex neural network-based decisions explainable, thereby providing methods to highlight which segments and sensors of a device contribute 
to a predicted anomaly \cite{audibert2020usad}. 
In contrast, we propose an \emph{intrinsically} interpretable white-box model that consists of a reduced set of interpretable features (or patterns). This enables oversight from human experts that can inspect patterns to trust decisions both case-by-case and globally. 
\section{Conclusions}
\label{conclusion}
In this paper, we present a method for detecting anomalies in networks of connected devices. To the best of our knowledge, no existing method is capable of dealing with this problem. In the first step, we propose a novel algorithm for detecting similar time series (of potentially different types of sensors) in different devices. In this way, we identify how devices interact with each other, as this information is often dynamic or unavailable. Once we have built a network of devices, we discover frequent cohesive patterns in the time series to describe normal device behaviour. To improve comprehension, we select the best patterns using the minimum description length principle. We then transform the time series into a pattern-based embedding, in which we search for anomalies. Intuitively, we see an anomaly as a time period in which frequent patterns are absent. By taking interactions between time series on the same device and various connected devices into account, we can capture contextual anomalies that would otherwise potentially be missed. In addition to being able to handle time series from a network of connected devices, our experiments show that our anomaly detection method also outperforms existing methods on both univariate and multivariate benchmark datasets.

\section*{Acknowledgements}
The authors would like to thank VLAIO CONSCIOUS (Anomaly Detection for Complex Industrial Assets) project for funding this research.

\bibliographystyle{plain}
\bibliography{references} 

\end{document}